\documentclass[twocolumn,aps,prl,amsmath,amssymb]{revtex4}

\usepackage{graphicx}

\usepackage{dcolumn}

\usepackage{bm}
\usepackage{amsmath,tabularx} 

\usepackage{epstopdf}

\DeclareGraphicsExtensions{.pdf,.eps,.png,.jpg,.mps}

\begin{document}

\preprint{AIP/123-QED}

\title{Visible and infrared photocurrent enhancement in a graphene-silicon Schottky photodetector through surface-states and electric field engineering}

\author{N. Unsuree$^{1}$}

\author{H. Selvi$^{1}$}

\author{M. Crabb$^{1}$}

\author{J.A. Alanis$^{2,3}$}

\author{P. Parkinson$^{2,3}$}

\author{T.J. Echtermeyer$^{1,3,4}$}

\email{tim.echtermeyer@manchester.ac.uk}

\affiliation{$^1$School of Electrical \& Electronic Engineering, University of Manchester, Manchester M13 9PL, UK}

\affiliation{$^2$School of Physics and Astronomy, University of Manchester, Manchester M13 9PL, UK}

\affiliation{$^3$Photon Science Institute, University of Manchester, Manchester M13 9PL, UK}

\affiliation{$^4$National Graphene Institute, University of Manchester, Manchester M13 9PL, UK}

\begin{abstract}

\noindent The design of efficient graphene-silicon (GSi) Schottky junction photodetectors requires detailed understanding of the spatial origin of the photoresponse. Scanning-photocurrent-microscopy (SPM) studies have been carried out in the visible wavelengths regions only, in which the response due to silicon is dominant. Here we present comparative SPM studies in the visible ($\lambda$ = 633nm) and infrared ($\lambda$ = 1550nm) wavelength regions for a number of GSi Schottky junction photodetector architectures, revealing the photoresponse mechanisms for silicon and graphene dominated responses, respectively, and demonstrating the influence of electrostatics on the device performance. Local electric field enhancement at the graphene edges leads to a more than ten-fold increased photoresponse compared to the bulk of the graphene-silicon junction. Intentional design and patterning of such graphene edges is demonstrated as an efficient strategy to increase the overall photoresponse of the devices. Complementary simulations and modeling illuminate observed effects and highlight the importance of considering graphene's shape and pattern and device geometry in the device design.

\end{abstract}

\maketitle

Graphene-silicon Schottky junctions have been extensively studied and their suitability as photodetectors (GSi-PDs) for radiation spanning the UV to the infrared regions has been demonstrated~\cite{DiBartolomeo2016,Wan2017,An2013,Liu2014,Srisonphan2016,Casalino2017,Selvi2018towards,selvi2018SOI,Amirmazlaghani2013,Wang2013,Goykhman2016}. The hybrid nature of GSi-PDs in which silicon forms the optical absorber for wavelengths $\lambda <$ 1100nm and graphene the light absorbing element for wavelengths $\lambda >$ 1100nm leads to high responsivities $R$ of devices of several 100mA/W in the visible and near-infrared wavelength regions but greatly reduced responsivities $R \ll$ 1mA/W in the infrared wavelength regime~\cite{Riazimehr2016}. The mechanisms responsible for the reduced efficiency of devices are both the reduced overall optical absorption A$_\text{gr}$ of graphene (A$_\text{gr}$ $\approx$ 2.3\%)~\cite{Nair2008} compared to bulk silicon, and the requirement for the optically excited charge carrier to overcome the Schottky barrier $\phi_b$ formed at the GSi interface by internal photoemission. The energy of each optically excited charge carriers $E_{exc}$ in graphene equals half the photon energy $E_{light}$, and needs to be on the order of the Schottky barrier height $\phi_b$ to significantly contribute to a measurable photocurrent~\cite{massicotte2016photo}. Scanning-photocurrent-microscopy (SPM) of GSi-PDs in the visible wavelength ranges has been used by Riazimehr, et al. to demonstrate the influence of parallel Schottky-MOS system formed in such devices and the drift of photo-excited carriers from underneath the MOS system to the Schottky region, leading to enhanced responsivities~\cite{riazimehr2017high,riazimehr2018high}. However, such SPM measurements are so far lacking for the infrared wavelength regime ($\lambda >$ 1100nm) which would allow a similar understanding of device operation for infrared photodetection.

\begin{figure*}[htbp]
\centering{
\includegraphics[width=170mm]{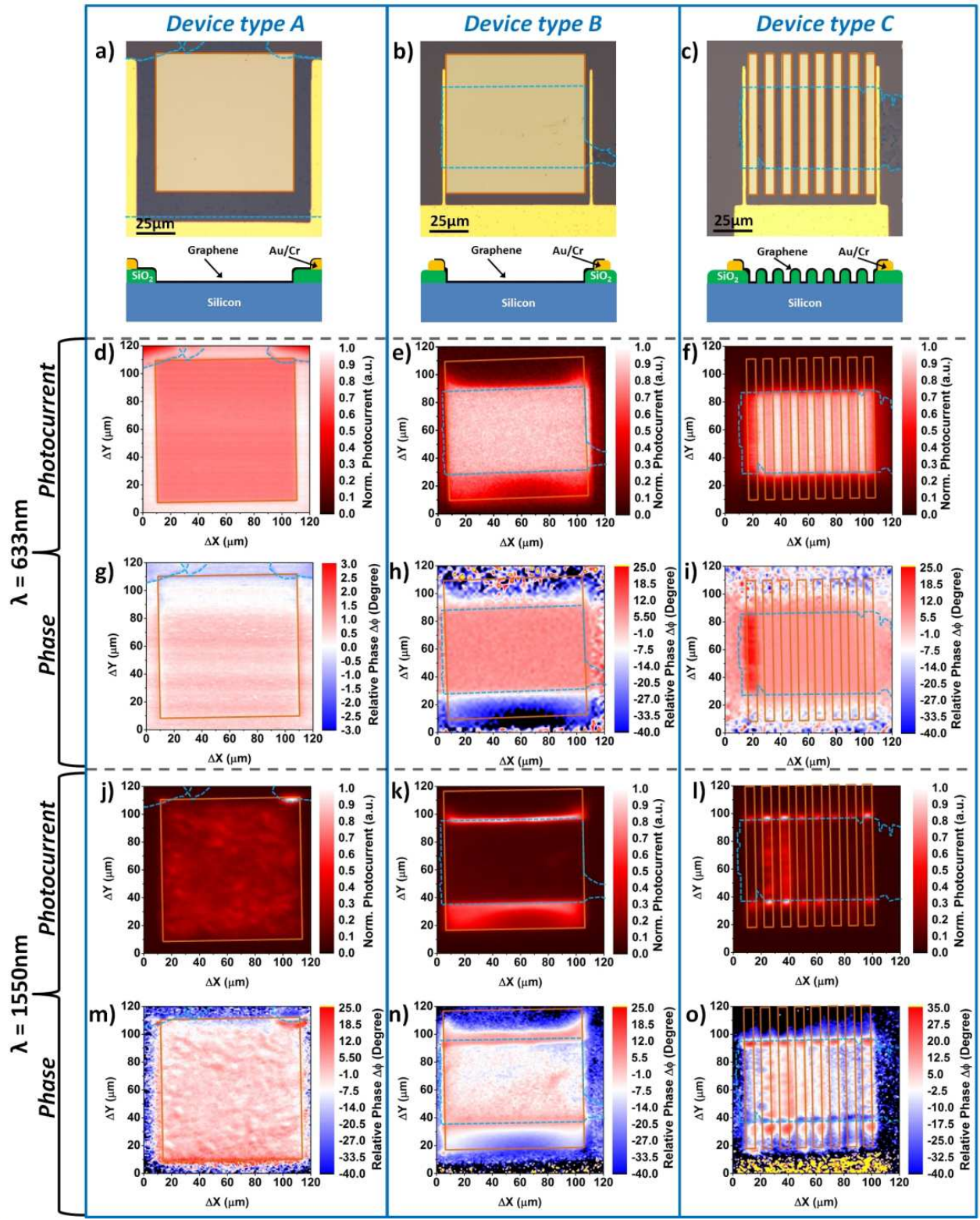}}
\caption{Optical micrographs (OM), schematic cross-sections and scanning photocurrent maps of three different device types. a) OM of device type A: graphene fully covers the oxide opening. b) OM of device type B: patterned graphene partially covers the oxide opening. c) OM of device type C: patterned graphene partially covers trenches opened in the oxide. d--f, g--i) Photocurrents maps for $\lambda$ = 633nm: normalized magnitude (d--f) and relative phase (g--i). j--l, m--o) Photocurrents maps for $\lambda$ = 1550nm: normalized magnitude (j--l) and relative phase (m--o). Indicated are the outline of graphene (dashed blue line) and the outline of the windows etched into the SiO$_2$ layer (orange line).}
\label{fig:SPM}
\end{figure*}

For our SPM studies, various GSi-PDs were fabricated employing processes described in more detail in~\cite{selvi2018SOI} and the SI. Figure \ref{fig:SPM}a--c shows optical micrographs of three different device types A, B, and C. As a starting substrate, low doped n-type silicon (Si) substrate with a doping level N$_d$ = 1$\times10^{15}$cm$^{-3}$ covered with a silicon dioxide (SiO$_2$) layer of thickness t = 100nm grown by thermal oxidation was used for all devices. Windows measuring 100x100 $\mu$m$^2$ were opened in the SiO$_2$ layer by wet chemical hydrofluoric acid (HF) etching for device types A and B. For device type C, arrays were opened in the SiO$_2$ layer by HF etching resulting in opened Si trenches of width w$_{\text{Si}}$ $\approx$ 6$\mu$m, interdigitated with SiO$_2$ covered Si ridges of width w$_{\text{SiO$_2$}}$ $\approx$ 6$\mu$m. Metal contacts consisting of chromium(Cr)/gold(Au) (thicknesses $t_{\text{Cr}}$ = 3nm, $t_{\text{Au}}$ = 40nm) were defined on the left and right hand side of the opened SiO$_2$ areas for all devices that lead into a large metal contact pad at the bottom of the window area. Chemical vapor deposition (CVD) grown graphene was transferred onto the substrates and patterned by an additional lithography and oxygen (O$_2$) plasma etching step. In device type A, the graphene layer completely covers the opened window, with the graphene-silicon Schottky junction formed over the whole opening (Figure \ref{fig:SPM}a). For device type B, the graphene layer is patterned such that the vertical dimension of the graphene sheet is smaller than the opened window and graphene's edges are in contact with the silicon substrate (Figure \ref{fig:SPM}b). Similarly, for device type C, the graphene sheet was defined smaller than the vertical length of the opened trenches and graphene's edges are in contact with the silicon substrate (Figure \ref{fig:SPM}c). These device architectures lead to the primary formation of a GSi Schottky junction for device types A and B, while in device type C the alternating Si and Si/SiO$_2$ regions form alternating Schottky and graphene-oxide-semiconductor (GOS) junctions~\cite{riazimehr2017high,riazimehr2018high}.

SPM measurements were conducted with laser light sources at wavelengths $\lambda$ = 633nm and 1550nm at laser intensities I $\approx$ 3$\mu\text{W}$/$\mu\text{m}^2$ and 125$\mu\text{W}$/$\mu\text{m}^2$ incident on the samples, respectively, which ensures similar absorbed optical fluence in silicon and graphene based on their optical absorption of A$_\text{silicon}$ = 100\% and A$_\text{graphene}$ = 2.3\%. Diffraction limited spot sizes were $\approx$ 1$\mu$m and 2$\mu$m, respectively, and SPM measurements were carried out at an applied reverse bias voltage of V$_b$ = -2V. The laser light was chopped at a frequency f = 2kHz and the electrical signals recorded using a lock-in amplifier (Zurich Instruments HF2LI/HF2TA)~\cite{Selvi2018towards}. Employed lock-in technique allows characterization of the photocurrent-magnitude and -phase to determine the absolute photoresponse as well as inferring time delays between optical excitation and electrical response of the devices. In a first step, all device types were characterized by SPM at a wavelength of $\lambda$ = 633nm. Shown in Figure \ref{fig:SPM}d--f are the magnitudes of the spatially dependent photocurrents. The magnitudes in the photocurrent maps have been normalized to their respective maxima for all devices to enable a comparison between all device types. In device type A, photocurrent is observed in the window region where the GSi Schottky junction is formed, as well as in the GOS regions formed by graphene on top of SiO$_2$ outside the opened area. The higher photocurrents in the GOS regions compared to the GSi Schottky junction can be attributed to the unintentional anti-reflection layer formed by the SiO$_2$ layer and the inversion layer formed in the GOS region itself~\cite{riazimehr2017high,riazimehr2018high}. The SiO$_2$ layer of thickness t = 100nm forms a $\lambda$/4 optical cavity with a reflection minimum at $\lambda \approx$ 633nm, matching the employed incident laser light wavelength which allows enhanced coupling of light with the silicon substrate (SI). In device type B, a photocurrent is not only generated at the GSi Schottky junction and the GOS region near the right hand side metal contact but also in the plain silicon regions above and below the GSi junction edge. Furthermore, a photoresponse can be observed in the SiO$_2$ $+$ Si regions to the left and right hand side of the metal contacts where graphene is absent. In device type C, the influence of the anti-reflective layer formed by the SiO$_2$ $+$ Si regions can again be observed, resulting in higher photocurrents in these regions of the device. A general feature observed for all device types in the visible wavelength region ($\lambda$ = 633nm) is the homogenous photocurrent generation in all graphene covered regions. This homogenous photoresponse can be attributed to silicon being the dominant optical absorber in the visible/near-infrared (NIR) wavelength regions at wavelengths $\lambda <$ 1100nm. Silicon's high optical absorption in combination with drift processes of photo-excited charge carriers over length scales of several $\mu$m in low doped silicon substrate allowing carriers to reach the GSi junction region lead to an efficient, homogenous conversion of photons into electrical currents~\cite{riazimehr2017high,riazimehr2018high}. Phase maps of the generated photocurrents confirm the homogenity of the photoresponses in the visible wavelength regions. Figure \ref{fig:SPM}g--i shows the relative phase maps $\Delta\phi$ for the different device types. The phase has been set to $\Delta\phi$ = 0$^\circ$ at a point in the center of the GSi junction for all device types. In device type A, the phase is almost constant over the whole device area. The minor phase change $\Delta\phi$ $\approx$ 1$^\circ$ of the photoresponse of optically excited charge carriers in the top compared to the bottom of the device equates to a time delay t$_\text{delay}$ $\approx$ 1$\mu$s based on the chopping frequency f$_\text{chop}$ = 2kHz. In device type B, the relative phase $\Delta\phi$ is again constant over the GSi junction area. In the plain silicon regions above (below) the top (bottom) graphene edge the relative phase $\Delta\phi$ resembles a similar shape to the magnitude of the photocurrent (Figure \ref{fig:SPM}e). Similar to device type A, device type C exhibits a homogenous, constant phase change $\Delta\phi$ in both the GSi and GOS junctions, resulting in an indistinguishable time delay of the photoresponse in the GSi and GOS regions except for a seemingly defective region near the left hand side metal contact.

The nature of photocurrent generation is different in the infrared wavelength regime. Figure \ref{fig:SPM}j--l shows SPM maps of the same devices acquired at a wavelength of $\lambda$ = 1550nm. In device type A, photocurrent generation occurs at the GSi Schottky junction only, i.e. the opened window area in which graphene is in contact with silicon. Photocurrent generation in the GSi junction region is consistent with the underlying physical model used to explain photocurrent generation in GSi Schottky junctions in the infrared wavelength regions~\cite{Sze2007,peters1967infrared,casalino2016internal}. Absorption of light in graphene and internal photoemission of charge carriers over the formed Schottky barrier leads to generation of photocurrents in GSi Schottky junction regions~\cite{massicotte2016photo}. In contrast with SPM measurements in the visible wavelength region (Figure \ref{fig:SPM}d) it can be observed that generated photocurrents are inhomogenous over the GSi junction area. This inhomogeneity can be attributed to the non-uniform interface and quality of the GSi Schottky junction. Previous studies highlighted the importance and implications of the natural oxide layer formed between graphene and silicon~\cite{Li2016a,Song2015} and the re-growth of this oxide layer after device fabriation~\cite{Selvi2018towards}. The interfacial oxide layer thickness and quality influence the Schottky barrier height and with it the efficiency of conversion of infrared light into electrical current. As such, SPM measurements in the infrared wavelength region present a powerful tool to study the spatially dependent properties of GSi junctions and highlight the need for further improvements and optimization of the GSi interface which is beyond the scope of the present study. Device type B presents a stark difference compared to device A in SPM measurements in the infrared wavelength regime. A strongly enhanced ($\approx$ 13 times) photocurrent contribution can be observed at the edges of the graphene sheet in contact with silicon compared to the bulk of the graphene sheet. Similarly, edge-enhanced infrared light detection can be observed in the top right corner of device type A where a defect during fabrication led to an unintentionally created graphene edge and at the graphene edges in device type C. Consistently with device type A, in device type C only the GSi junctions formed in the opened trenches of the array lead to a photoresponse while no response can be observed in the GOS junctions. It can be further observed that the infrared photoresponse varies significantly across the GSi junctions formed in the centers of the different trenches. We attribute this inhomogenity to the imperfect conformal transfer of graphene onto the 3-dimensionally patterned substrate. The phase map for device type A reveals an inhomogenous relative phase change $\Delta\phi$ across the GSi junction area whose features resemble similarities to the features observed for the magnitude of the photoresponse (Figure \ref{fig:SPM}j). In device type B, an inhomogenous phase change is again observed in the center of the GSi junction. However, the relative average phase difference $\Delta\phi_\text{edge}$-$\Delta\phi_\text{center}$ $\approx$ 10$^{\circ}$ between the edges and the center of the GSi junction indicates an increase in photoresponse speed of the graphene edges compare to the center of $t_\text{resp,edge}$-$t_\text{resp,center}$ $\approx$ 14$\mu$s. The relative phase $\Delta\phi$ in device type C demonstrates a speed enhancement at the graphene edges and indicates that an increased photoresponse magnitude (Figure \ref{fig:SPM}l) corresponds to increased response speed.

\begin{figure}[htbp]
\centering{
\includegraphics[width=80mm]{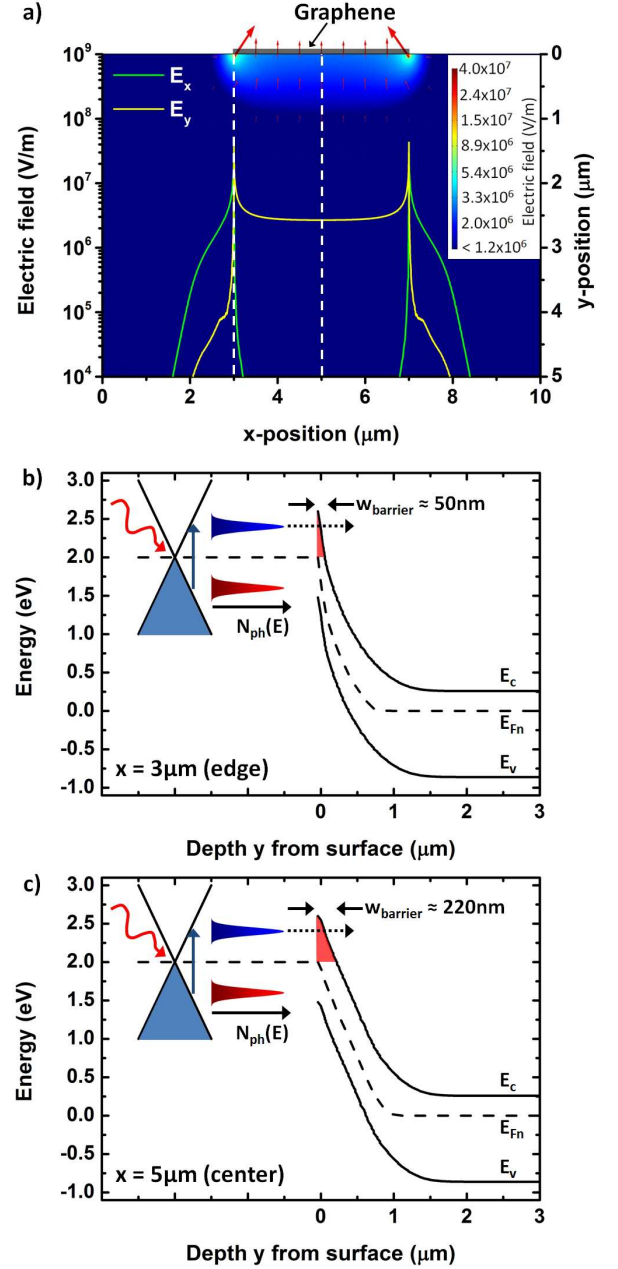}}
\caption{Simulation of a GSi Schottky junction under a reverse bias of V$_b$ = -2V. a) 2-dimensional plot of the electric field strength and direction in the silicon substrate (color coded in logarithmic scale and arrows (direction), respectively). Line-plots: Electric field components E$_x$ and E$_y$ 10nm below the silicon surface along the x-direction. b,c) Energy band diagrams of the graphene-silicon Schottky junction as a function of depth y at positions x = 3$\mu$m and 5$\mu$m at the edge and the center of the Schottky junction, respectively (vertical white dashed lines in a). Qualitatively indicated is the non-equilibrium carrier density (N$_{ph}$) in the conduction and valence bands in graphene after optical excitation at a wavelength of $\lambda$ = 1550nm (E$_{ph}$ = 0.8eV) before thermalization.}
\label{fig:efieldbands}
\end{figure}

Finite-element-method simulations (COMSOL) have been carried out to further understand the observed edge-enhanced photocurrents at infrared wavelengths. An ideal GSi Schottky junction has been modeled in 2-dimensions in which a strip of graphene of width 4$\mu$m and workfunction $\Phi_G$ = 4.6eV~\cite{Yu2009} has been placed on top of a silicon substrate of height h = 5$\mu$m and width w = 10$\mu$m with a doping concentration N$_d$ = 1$\times$10$^{15}$cm$^{-3}$ under a reverse bias of V$_b$ = -2V. Figure \ref{fig:efieldbands}a shows a 2-dimensional plot of the electric field strength (color) and direction (arrows) inside the silicon substrate. Consistent with well-studied electrostatic effects at discontinuities in metal-semiconductor contacts, the electric field is strongly enhanced at the edges of the graphene strip~\cite{Sze2007}. A line-plot of the electric field components E$_x$ and E$_y$ along the x-direction 10nm below the silicon surface demonstrates an enhancement of the y-component of the electric field E$_y$ of approximately one order of magnitude at the edges of the graphene strip compared to its center. While the y-component of the electric field E$_y$ quickly decays outside the graphene strip, the electric field component E$_x$ decays less rapidly, resulting in an in-plane electric field outside the graphene strip. The center of the graphene strip is dominated by an almost constant electric field component E$_y$ and the electric field component E$_x$ is negligible. The difference in electric field strengths at the edges and in the centre of the graphene strip is further manifested in the energy band diagrams. Figure \ref{fig:efieldbands}b,c show the simulated energy band diagrams as a function of depth from the surface of the silicon substrate at the edge and in the center of the GSi junction, indicated by the white dashed lines in Figure \ref{fig:efieldbands}a. Also indicated is the creation of the non-equilibrium charge carrier density (N$_{\text{ph}}$) under optical excitation at a wavelength $\lambda$ = 1550nm (E$_{ph}$ = 0.8eV). Increased band bending at the graphene edge (Figure \ref{fig:efieldbands}b) compared to the center (Figure \ref{fig:efieldbands}c) results in a reduced effective Schottky barrier width and height at the edge~\cite{Sze2007}. The barrier widths at the Fermi level are w$_{\text{barrier}}$ $\approx$ 50nm and 220nm at the graphene edge and in the center, respectively. Assuming a triangular potential barrier, optically excited charge carriers encounter an even further reduced effective barrier width at an energy of 0.4 eV above the Fermi level (E$_{ph}$/2) of w$_{\text{barrier}}$ $\approx$ 17nm and 73nm at the graphene edge and in the center, respectively. The simulated band energy diagrams and corresponding electric field enhancement allow to qualitatively deduce that strong band bending in the silicon substrate at the graphene edges will profoundly influence the transmission of photoexcited charge carriers in graphene over/through the graphene-silicon Schottky barrier. Resulting, photoexcited charge carriers, particular with energies below the Schottky barrier height $\Phi_b$, can be extracted more efficiently at the edges of the graphene sheet compared to the center as observed in our experiments.

\begin{figure}[htbp]
\centering{
\includegraphics[width=80mm]{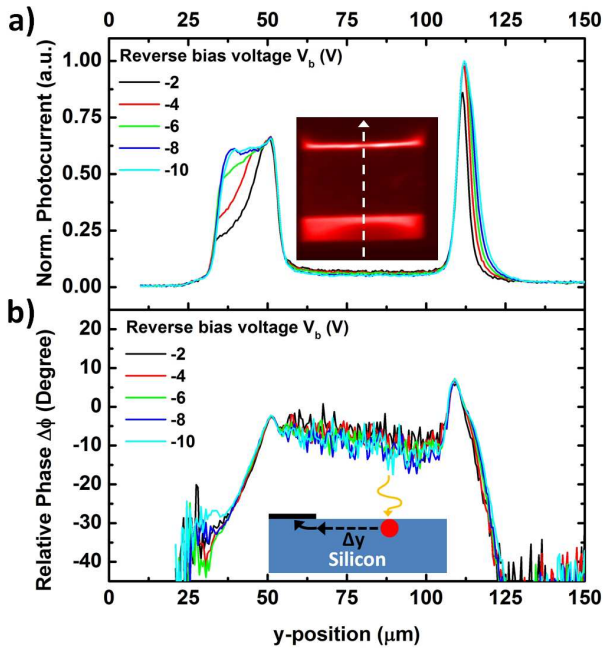}}
\caption{Vertical linescans of the photocurrent magnitude (a) and phase (b) across the center of device type B, indicated by the white arrow in the inset in (a). The charge carrier transport mechanism is schematically indicated in the inset in (b).}
\label{fig:line}
\end{figure}

The graphene edge-enhanced photocurrents in device type B have further been studied at a wavelength $\lambda$ = 1550nm by vertical linescans across the center of the device as a function of applied reverse bias voltage V$_b$. Figure \ref{fig:line}a) shows that the over one order of magnitude enhanced maximum photocurrent generated at the upper and lower graphene edges increases only weakly with increasing reverse bias voltage. However, surprisingly the photocurrent peaks formed at the upper and lower edges of the device differ significantly in shape and as a function of applied reverse bias voltage V$_b$. A narrow peak of full-width-half-maximum (FWHM) $\approx $ 4$\mu$m can be observed at the top graphene edge that increases to a FWHM $\approx $ 7$\mu$m for V$_b$ = -2V and V$_b$ = -10V, respectively. On the contrary, the bottom graphene edge exhibits an almost three-fold increased photocurrent peak width compared to the photoresponse at the top edge with a FWHM of $\approx $ 11$\mu$m and $\approx $ 19$\mu$m, for V$_b$ = -2V and V$_b$ = -10V, respectively, that significantly extends spatially into the silicon areas not covered by graphene. Intriguingly, we see great similarity between the photocurrent patterns at visible and infrared wavelengths, shown in Figures \ref{fig:SPM}e \& k, respectively. The spatial photoresponses of the device in the plain silicon regions above (below) the top (bottom) graphene edges bear striking resemblances for optical excitation with both visible and infrared light, e.g. observed curvatures and patterns of the photoresponses in these regions are identical. We attribute the photoresponse observed in the plain silicon regions at infrared wavelengths to optical absorption by surface states in the exposed silicon surface~\cite{chiarotti1971optical,casalino2010near}. Charge carriers trapped in surface states with sub-band gap energies E$_\text{ss}$ can be freed by infrared light and drift into the junction area due to the in-plane electric field component E$_\text{in-plane}$ present in the silicon substrate. The asymmetric device structure and corresponding electrostatics due to the presence of metal fingers on the left and right hand side of the window area and the large metal contact pad at the bottom of the device (Figure \ref{fig:SPM}b) lead to a spatially enlarged photoresponsive silicon area below the bottom graphene edge compared to the region above the top graphene edge. The MOS-systems formed by the metal contacts on top of the oxide on the bottom/side of the device are in the inversion regime at a reverse bias voltage of V$_b$ = -2V as confirmed by simulations. These MOS-systems contribute in parallel with the GSi junction to the formation of the in-plane electric field component E$_\text{in-plane}$ with increased spatial extent in the bottom region of the device compared to the upper device region (Figure \ref{fig:line}a). Possible thermal effects, e.g. heating of silicon with infrared light can be ruled out due to the reverse bias voltage dependence of the spatial extend of the photocurrent generation area which contradicts thermal transport effects that are electric field independent in silicon. Charge carrier diffusion processes can be excluded as these also are voltage independent and further would result in equal photoresponse features on the top and bottom sides of the graphene edges. The conclusion of surface states being responsible for optical absorption in silicon at wavelength $\lambda >$ 1100nm is further confirmed by the observation that the photoresponse abruptly disappears in the silicon regions in the bottom of the device covered by SiO$_2$ (Figure \ref{fig:SPM}k). The sharp switch-off of the photocurrent at the Si-Si/SiO$_2$ interface is consistent with the passivation of surface states in SiO$_2$ covered silicon~\cite{Tyagi1984,chiarotti1971optical,Card1977,Song2015}. The reverse bias voltage dependent phase of the photoresponse allows determining the spatially dependent response times of charge carriers generated in the plain silicon regions. Figure \ref{fig:line}b shows that above (below) the top (bottom) graphene edge an abrupt, approximately linear phase change $\Delta\phi$ occurs beyond the graphene edges, indicating that charge carriers contributing to the photoresponse in plain silicon exhibit a linearly increasing time delay the further away from the edges they are excited. The inset in Figure \ref{fig:line}b schematically depicts our proposed underlying photoresponse mechanism, based on the Haynes-Shockley experiment~\cite{haynes1949investigation}, in which carriers excited in silicon outside the graphene edge need to travel to the GSi Schottky junction before contributing to the photocurrent. The spatially dependent phase delay with respect to the graphene edges

\begin{equation}
\Delta\phi_\text{edge}(\Delta\text{y}_\text{edge}) = \Delta\phi(\Delta\text{y}_\text{edge})-\Delta\phi(\Delta\text{y}=y_\text{edge})
\label{eq:relphi}
\end{equation}

can be used to approximate the position dependent response time 

\begin{equation}
\text{t}_{\text{resp}}(\Delta\text{y}_\text{edge}) \approx \frac{\Delta\phi_\text{edge}(\Delta\text{y}_\text{edge})}{360^\circ} \times \frac{1}{f_\text{chop}}    
\label{eq:reltime}
\end{equation}

From a linear fit of the position dependent response time t$_{\text{resp}}(\Delta\text{y}_\text{edge})$ the average carrier drift velocity $\text{v}_\text{drift}$ in the top (bottom) silicon region above (below) the graphene edge can be calculated to $\text{v}_\text{drift}(\Delta\text{y}_\text{edge})$ = 18m/s and 35m/s, respectively, with a minor dependence on the reverse bias voltage. Assuming a hole mobility $\mu \approx$ 100$\frac{\text{cm}^2}{\text{Vs}}$, the general relation between carrier drift velocity v$_\text{drift}$ and electric field E, v$_\text{drift}$ = $\mu$E, yields an effective in-plane electric field E$_\text{in-plane} \approx$ 18V/m and 35V/m in the top and bottom silicon regions, respectively. The slightly lower in-plane electric field E$_\text{in-plane}$ in the top compared to the bottom silicon area of the device is in good qualitative agreement with the spatially narrower photoresponse at the top graphene edge (Figure \ref{fig:line}a).

\begin{figure*}[htbp]
\centering{
\includegraphics[width=170mm]{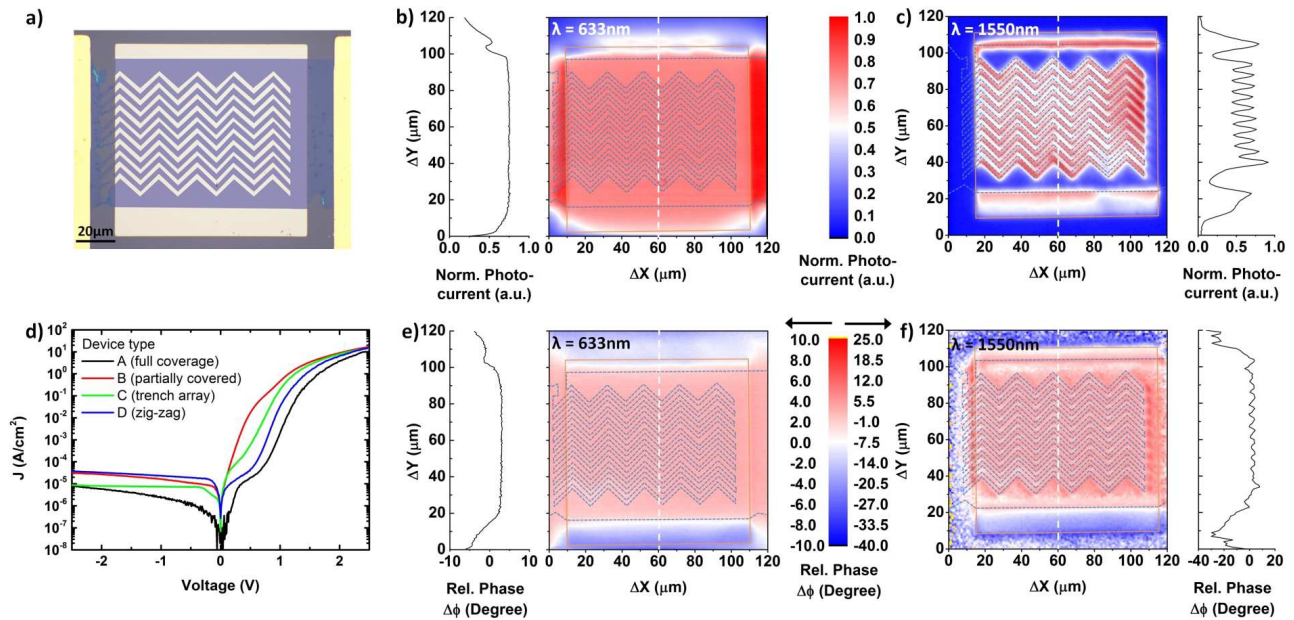}}
\caption{Engineered GSi-PD (device type D) with increased perimeter of graphene edges. a) Optical micrograph with overlaid graphene outline. b,c) Photocurrent magnitude maps and vertical line scan in the center of the device (indicated by dashed white line) at wavelengths of $\lambda$ = 633nm and 1550nm, respectively. d) JV characteristics of all device types. e,f) Photocurrent phase maps  and vertical line scan in the center of the device (indicated by dashed white line) at wavelengths of $\lambda$ = 633nm and 1550nm, respectively.}
\label{fig:zigzag}
\end{figure*}

We have exploited the observed edge-enhancement to fabricate a device (type D) with additional, intentionally created graphene edges to further systematically enhance infrared light detection of GSi-PDs. Figure \ref{fig:zigzag}a shows the optical micrograph of device type D in which the perimeter of the graphene edges has been increased by patterning graphene into zigzag lines by a lithography and O$_2$ plasma etching step. SPM measurements carried out at visible wavelength ($\lambda$ = 633nm) show the previously observed pattern of homogenous photocurrent generation across the GSi junction. Both magnitude and phase of the photocurrent do not reveal the defined zigzag pattern (Figure \ref{fig:zigzag}b,e). However, in the infrared wavelength regime at $\lambda$ = 1550nm defined zigzag pattern is clearly distinguishable in both magnitude and phase of the photocurrent maps (Figure \ref{fig:zigzag}c,f). The patterned edges in device type D lead to a significantly increased photocurrent generation in magnitude and increase the photoactive area. An overall 5 times enhanced photocurrent of the device can be derived from the ratio of the area-averaged photocurrents generated in the center of the device with zigzag pattern to the photocurrent generated in the bulk of the unpatterned graphene in the blue triangular areas above/below the zigzag pattern (Figure \ref{fig:zigzag}c). This 5 times enhancement is in good agreement with experimental photocurrent characterization of device types A and D under global illumination at $\lambda$ = 1550nm, i.e. when light is incident on the whole device area as opposed to SPM measurements. It is further noteworthy that graphene patterning and corresponding graphene perimeter increase do not have a detrimental impact on the current-density--voltage (JV) characteristics of the devices. Figure \ref{fig:zigzag}d shows the JV curves for all devices types A-D and demonstrates that all device types exhibit and on- to off-current ratio I$_\text{on}$/I$_\text{off}$ greater than 5 orders of magnitude. Patterned devices exhibit approximately 3 times increased off-current densities compared to the un-patterned device type A under a reverse bias voltage V$_b$ = -2.5V, attributable to the local electric-field hotspots at the edges of the graphene sheet, and on-current densities comparable to the un-patterned device type A under a forward bias voltage V$_b$ = +2.5V. Yet, despite the increased off-current density of the patterned devices compared to the un-patterned device, patterning of devices has not been found detrimental for photodetector applications. All four device types exhibit a similar noise current level with a standard deviation $\sigma$ = 1.6pA derived from time-dependent current-voltage (IV) measurements in the dark under an applied reverse bias voltage V$_b$ = -2V at a sample rate f$_s$ = 7Hz. Further observable differences in the JV characteristics such as slopes and thresholds are currently beyond the scope of this study.

In conclusion, our SPM characterization of various GSi-PDs device architectures reveals the significant difference in the photocurrent generation mechanism of GSi-PDs in the visible and infrared wavelength range, respectively, and its manifestation in the spatial and time domain. We demonstrate electric-field enhancement by graphene-perimeter- and device-electrostatics-engineering as an efficient method to increase the photoresponse in GSi-PDs in the infrared at wavelengths $\lambda >$ 1100nm by more than one order of magnitude. Simulations reveal that local electric-field hotspots corresponding to large gradients of the energy bands below the surface of the silicon substrate form at the graphene edges which lead to a more efficient conversion of optically excited charge carriers into electrical currents in the infrared wavelength regime. Besides spatially resolved photocurrent magnitudes, we show that phase maps can provide additional invaluable information about local photoresponse times. Unexpectedly, we discover that surface states in the silicon substrate can lead to an additional photoresponse component in GSi-PDs at infrared wavelengths with response times in the $\mu$s range. Overall, SPM at wavelengths $\lambda >$ 1100nm is an invaluable tool to study GSi-PDs and probe spatial responses, time delays and interfacial inhomogeneities. In the future, we expect further significant GSi-PD enhancement based on surface state engineering, graphene perimeter-to-area and electro-static optimization of devices and dopant profile engineering in the silicon substrate in the form of e.g. guard-rings and intentionally designed tunneling barriers~\cite{Sze2007}. The observed edge-enhancement potentially forms a foundation for novel devices such as proximity-coupling of GSi-PDs with plasmonic metal structures, introducing graphene-nano-ribbons (GNRs) into the GSi device architecture as well as enhanced photochemical sensing due to graphene's highly chemically reactive edge~\cite{bellunato2016chemistry}. We believe that demonstrated surface state effects and device engineering concepts will not only be beneficial for GSi-PDs but can also be extended towards other 2D-materials.

\section{Acknowledgements}

N.U. acknowledges funding from The Royal Thai Army. H.S. acknowledges funding from the Turkish government (MEB-YLSY). P.P. acknowledges funding from the Royal Society (RG140411). T.J.E acknowledges funding from the Huawei Innovation Research Program (HIRP).



\end{document}